# Taylor microscale and effective Reynolds number near the Sun from PSP

C. Phillips[1], R. Bandyopadhyay[1], D. J. McComas[1]

[1]Department of Astrophysical Sciences, Princeton University, Princeton, NJ 08544, USA

**Abstract**
The Taylor microscale is a fundamental length scale in turbulent fluids, representing the end of fluid properties and onset of dissipative processes. The Taylor microscale can also be used to evaluate the Reynolds number in classical turbulence theory. Although the solar wind is weakly collisional, it approximately behaves as a magnetohydrodynamic (MHD) fluid at scales larger than the kinetic scale. As a result, classical fluid turbulence theory and formalisms are often used to study turbulence in the MHD range. Therefore, a Taylor microscale can be used to estimate an effective Reynolds number in the solar wind. NASA's Parker Solar Probe (PSP) has reached progressively closer to the Sun than any other spacecraft before. The collected data have revealed many new findings in the near-Sun solar wind. Here, we use the PSP data to estimate the Taylor microscale and effective Reynolds number near the Sun. We find that the Taylor microscale and Reynolds number are small compared to the corresponding near-Earth values, indicating a solar wind that has been less processed by turbulence, with very small-scale dissipative processes near the Sun.

## 1. Introduction

In a turbulent system, energy is distributed over a broad range of length scales. In the classical cascade scenario, the energy is transferred from large scales to small scales by non-linear processes (Kolmogorov 1941). Eventually this cascade reaches small scales where dissipative processes become dominant and dissipate the fluid motion to random individual particle motions or heat. This generic model holds in hydrodynamic turbulence and magnetohydrodynamics (MHD) turbulence for sufficiently turbulent systems. Broadband fluctuations are observed in solar wind measurements, such as the magnetic field, velocity, and density (Coleman 1968; Matthaeus & Goldstein 1982). Understanding the spatio-temporal structure of solar wind turbulence is important for modeling transport of electromagnetic energy, mass, and energetic particles (Breech et al. 2008; Tessein et al. 2013 ; Bandyopadhyay et al. 2020c). Characteristic scales such as the Taylor microscale and correlation scale provide a foundation for this understanding by quantifying the turbulence in the system.

The correlation scale represents the characteristic size of the largest eddies in a turbulent system. The Taylor microscale represents the length of the smallest structures which are dominated by non-linear dynamics rather than dissipative forces and defines the lower boundary of the inertial range (Frisch 1995). The correlation length and Taylor scale can be used to evaluate the Reynolds number, which measures the level of turbulence. A large Reynolds number indicates that the strength of the non-linear processes is much greater than the strength of the dissipative processes at large scales. Taylor microscale has been long studied in hydrodynamic simulations and laboratory fluids (e.g., Pearson et al. 2002, 2004). However, in contrast to other characteristic length scales, Taylor microscale has been rarely explored in space plasma turbulence, mostly due to instrumental limitations. Taylor microscale measurements require high-resolution data, which are often infeasible. In contrast, the correlation scale is easily calculated from low-resolution data, allowing for strict upper limits on the size of eddies in the inertial range. The first measurement of Taylor scale and effective Reynolds number in the solar wind came from a study by Matthaeus et al. 2005. Subsequently, a series of works (Bandyopadhyay et al. 2020b; Chuychai et al. 2014; Gurgiolo et al. 2013; Matthaeus et al. 2008; Weygand et al. 2009, 2009, 2011) refined the methodology and reported measurements of Taylor microscale and Reynolds number in the solar wind and planetary magnetospheres. Recently, the Taylor microscale and Reynolds number have been measured in laboratory plasmas (Cartagena-Sanchez et al. 2022). The Taylor microscale, correlation scale, and Reynolds number depend on solar wind conditions (Zhou et al. 2020; Zhou & He 2021). Therefore, it is of interest to evaluate these quantities in the previously unexplored regions of the heliosphere near the Sun, where the solar wind originates.

Launched on August 12, 2018, NASA's Parker Solar Probe (PSP), even in its first orbit, reached closer to the Sun than any previous spacecraft (Fox et al. 2016). The initial orbits of PSP have already returned a treasure trove of scientific data revealing new insights into the nature of solar wind close to the Sun (e.g., Bale et al. 2019; Kasper et al. 2019; McComas et al. 2019). Nevertheless, quantitative measurement of Taylor scale and Reynolds number near the Sun from PSP data have not yet been yet reported (although see Cuesta et al. 2022). In this paper, we use the high-resolution magnetic field data from PSP to evaluate single-spacecraft measurements of Taylor scale in the solar wind close to the Sun. The quantitative estimation of Taylor microscale, in turn, permits an evaluation of an effective magnetic Reynolds number under the assumption of a standard resistive or Ohmic dissipation function.

## 2. Theory and Technique

Both Taylor microscale and correlation scale are related to the two-point correlation function, here defined for magnetic field, as

$$R(\tau) = \langle \boldsymbol{b}(t) \cdot \boldsymbol{b}(t+\tau) \rangle_T, \qquad (1)$$

where, $\boldsymbol{b}$ is the fluctuating magnetic field, and $\langle ... \rangle_T$ is a time average over the total time span of the data. Using Taylor's frozen-in hypothesis ($r = V_{SW}\,\tau$), the temporal correlation function ($R(\tau)$) can be converted to a spatial correlation function ($R(r) = \langle \boldsymbol{b}(\boldsymbol{x}) \cdot \boldsymbol{b}(\boldsymbol{x}+\boldsymbol{r}) \rangle$). The correlation function at zero lag ($\tau = 0$) is the variance ($R(0) = \langle b^2 \rangle$), and we normalize the correlation function by the variance.

The Taylor microscale (Eq. (2)) is the measure of the curvature of the correlation function ($R(r)$) at zero separation.

$$\lambda_T = \sqrt{-\frac{R(0)}{R''(0)}}\ . \qquad (2)$$

For statistically homogeneous systems, $R(r) = R(-r)$. Therefore, for small length scales $r$, the correlation function (Eq. (1)) can be expanded in Taylor series near the origin as

$$R(r) = 1 - \frac{r^2}{2\lambda_T^2} + \cdots, \qquad (3)$$

where we have assumed isotropy and the higher order terms are ignored.

The correlation length ($\lambda_C$) is related to the characteristic size of the energy-containing eddies in a turbulent system (Pope 2000). Here, we compute $\lambda_C$ from the two-point correlation function (Eq. (1)). We use the standard Blackman-Tukey method (Blackman & Tukey 1958; Matthaeus & Goldstein 1982), with subtraction of the local mean, to evaluate the correlation function from Eq. (1). We employ the "1/$e$" definition (e.g., Matthaeus & Goldstein 1982, Smith et al. 2001), namely

$$R(\lambda_C) = \frac{R(0)}{e}. \qquad (4)$$

The classical Reynolds number in a turbulent fluid is defined as

$$Re = \frac{v_{\text{rms}} \lambda_C}{\nu}, \qquad (5)$$

where $v_{\text{rms}}$ is the root-mean square velocity fluctuation and $\nu$ is the fluid viscosity. Eq. (5) can be reformulated as

$$Re = \left(\frac{\lambda_C}{\eta}\right)^{4/3}, \qquad (6)$$

where $\eta = \left(\frac{\nu^3}{\epsilon}\right)^{1/4}$ is the Kolmogorov length scale, which represents the "dissipation scale" where structures are critically damped, so that non-linear dynamics are negligible and dissipative processes are dominant. Here $\epsilon$ is the average rate of dissipation. In the solar wind, however, the viscosity $\nu$ cannot be determined because the solar wind plasma is very weakly-collisional. Some studies have attempted to estimate an effective Reynolds number by evaluating Eq. (6) with replacing $\eta$ by the ion-kinetic scales such as the ion-inertial length ($d_i$) or the ion gyro radius ($\rho_i$) (Parashar et al. 2015, 2019). However, in the solar wind all of the energy is not dissipated at the ion scales (Alexandrova et al. 2009; Sahraoui et al. 2009) and this assumption may not always be accurate. Therefore, we resort to another indirect of way of estimating the Reynolds number. From hydrodynamic turbulence theory (Batchelor 1953; Tennekes & Lumley 1972), we can calculate an effective Reynolds number from the measured correlation scale ($\lambda_C$) and Taylor microscale ($\lambda_T$).

$$Re = \left(\frac{\lambda_C}{\lambda_T}\right)^2. \qquad (7)$$

This method of estimating the Reynolds number bypasses the need to identify any specific dissipation scale, and therefore, is more generally applicable.

We note that the above formulae are derived from classical hydrodynamic theory but can be applied to a MHD fluid under the assumption of isotropy. The solar wind plasma is weakly collisional, but many MHD properties are observed at large length scales (e.g., Matthaeus & Goldstein 1982; Sorriso-Valvo et al. 2007). Therefore, often MHD theories and formalisms are applied to characterize solar wind fluctuations.

## 3. PSP Data

To compute the Taylor microscale, correlation length, and effective Reynolds number we use the SCaM data product of the FIELDS suite (Bale et al. 2016, 2019), which merges Fluxgate magnetometer (MAG) and search-coil magnetometer (SCM) measurements (Bowen et al. 2020). The SCaM dataset sufficiently resolves the ion-kinetic range well above the noise floor (signal-to-noise ratio > 5) (Huang et al. 2021). Due to a SCM anomaly in 2019 March, the full three-dimensional SCaM data are available only for part of the first solar encounter by PSP and are unavailable for all subsequent encounters. Therefore, our analysis is limited to 2018 November 4 and November 7 for a total of about two days and 10 hours ($\approx$ 58.5 hours) of data. We divide the data in this interval into 15-minute subintervals and calculate the Taylor microscale and Reynolds number in each of them. Radial distance from the Sun varies from a maximum of 38.03 solar radii ($R_s$) to a minimum of 35.7 $R_s$ during the data collection period, with an average of 36.3 $R_s$.

We obtain bulk plasma properties, including proton density ($n_p$) and velocity from the Solar Probe Cup (SPC) data in the SWEAP instrument suite (Case et al. 2020, 20; Kasper et al. 2016, 2019). The SPC plasma moments are averaged in every 15-min interval. We use the average plasma measurements to compute plasma variables, such as the Alfvén speed ($V_A = |\boldsymbol{B}|/\sqrt{\mu_o m_p n_p}$) for each interval. The ratio of average solar wind speed ($V_{SW}$) to the average Alfvén speed ($V_A$) remains greater than 2 for all the cases, so Taylor's frozen-in approximation (Taylor 1938) is roughly valid.

We note here that Eqs. (1)-(7) can be evaluated using velocity field or Elsasser field as well, and indeed such calculations using electron velocities have been conducted in the near-earth solar wind (Gurgiolo et al. 2013b). However, the PSP proton velocity field data are mostly low resolution, reaching the end of the inertial range (e.g., Chen et al. 2020). Even when higher resolution velocity data are available, the high-frequency part is likely affected by noise (Vech et al. 2020). Therefore, we refrain from using the velocity field and rely solely on the high cadence magnetic field data.

## 4. Results

For every 15-minute period we first calculate the magnetic correlation function from the PSP SCaM data. As an example, Fig. 1 shows the magnetic field, proton velocity, and the proton density data in a sub-interval covering the 15-minute period from 11:14:59 UTC to 11:29:59 UTC on 2018 November 6. The magnetic field and proton velocity are shown in the RTN coordinate system (Franz & Harper 2002). The R (radial) axis is defined by the radial vector from the Sun to the spacecraft, directed radially away from the Sun; the T (tangential) axis is in the direction of the cross product of the Sun's spin vector and the R axis, and the N (normal) axis completes the right-handed coordinate set, taken largely from Zieger et al. 2009.

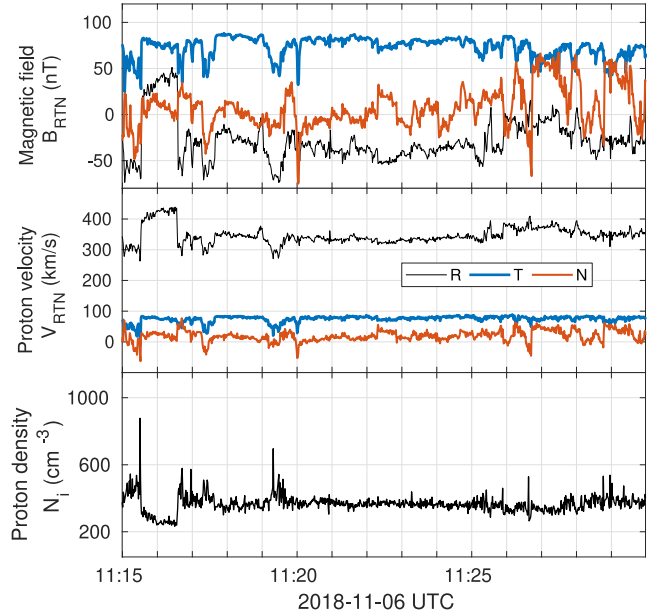

**Figure 1:** Overview of the PSP data in the 15-minute interval between 11:14:59 UTC and 11:29:59 UTC on 2018 November 6. The data are from the FIELDS and SWEAP instrument suites onboard the PSP spacecraft. The panels (from top to bottom) plot the magnetic field components in RTN coordinate, the proton velocity in RTN coordinates, and the proton number density.

In Fig. 2 we show the correlation function ($R(r)$), computed from the magnetic field data in this 15-minute period on 2018 November 6. The average bulk speed in this interval is $V_{SW} = 345$ km/s, and we use this value to convert temporal scales to spatial lags. From Eq. (3), we can estimate the Taylor microscale by fitting the correlation function $R(r)$ to a parabola ($R = 1 - r^2/2\lambda_T^2$) near the origin. Evidently, the quadratic approximation becomes more accurate as one asymptotically approaches smaller values of $r$, but the precision of the estimation increases with the maximum amount of lag $r$. Therefore, we estimate the Taylor microscale following the methods of Chuychai et al. (2014), which enable accurate and precise estimation. The technique can be summarized in the following way: The magnetic-field correlation function is computed for each 15-min interval. The Taylor scale is then estimated using a series of maximum lag approximations ranging from a minimum lag of 3 data points to a maximum lag of 20. Then, we fit a straight line to these estimated values of the Taylor scale as a function of maximum lag, and the intercept at lag = 0 is taken as the final estimated value of the Taylor microscale. The error is the standard deviation given by the fit.

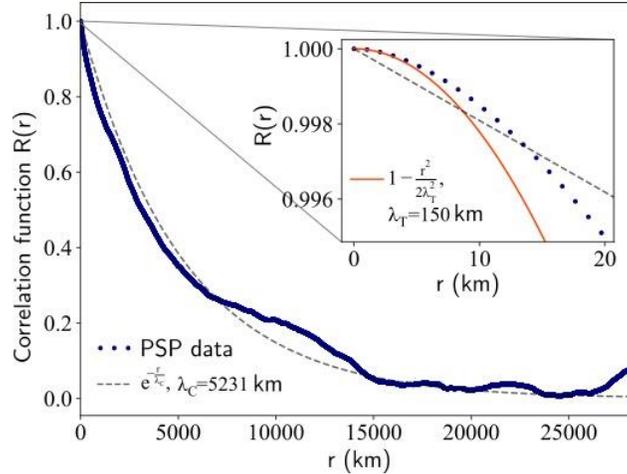

**Figure 2:** One example of magnetic correlation function from PSP data (blue dots) during the time interval from 11:14:59 UTC to 11:29:59 UTC on 2018 November 6. The exponential fit (gray, dashed line) is used to estimate the correlation length, and a parabolic fit near the origin (red, continuous line) is used to estimate Taylor scale. Inset: Computed correlation function, exponential fit, and parabolic fits for small length scales.

We show the PSP data-based evaluation of the correlation function using blue dots in Fig. 2. The quadratic fit at the origin, following the methods of Chuychai et al. (2014), is shown by the red curve in the enlarged inset. The quadratic curve only fits the points near the origin and begins to deviate at larger spatial scales (see Eq. 3). The resulting Taylor microscale value is $\lambda_T = 150 \pm 2$ km, where the uncertainty is the standard deviation given by the fit. We repeat the same procedure for all 15-minute intervals. The resulting Taylor scale values along with their respective uncertainties are shown in Fig. 3. The black horizontal line is the average, and the shaded region represents one standard deviation spread.

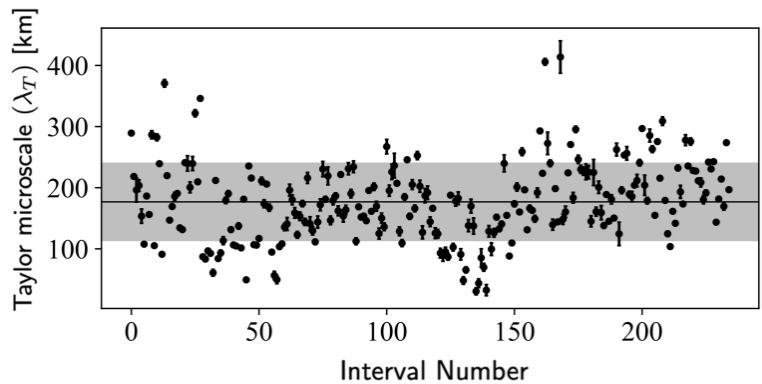

**Figure 3:** Calculated Taylor microscale for the PSP intervals in the first encounter. The average and standard deviation are shown by the black horizontal line and the shaded region.

From Fig. 2, at large spatial scales the correlation function exhibits an approximate exponential behavior, and an exponential fitting (see Eq. (4)) provides a reasonable estimate of the correlation length (Smith et al. 2001). Here, the estimated value of correlation length is $\lambda_C = 5231 \pm 7$ km, consistent with previous reports (Bandyopadhyay et al. 2020; Chen et al. 2020; Parashar et al. 2020, Cuesta et al. 2022). The error is estimated by the standard deviation between the fit and the computed correlation function points. Like the Taylor microscale, we calculate the correlation length for every fifteen-minute interval. The correlation length estimates are plotted in Fig. 4.

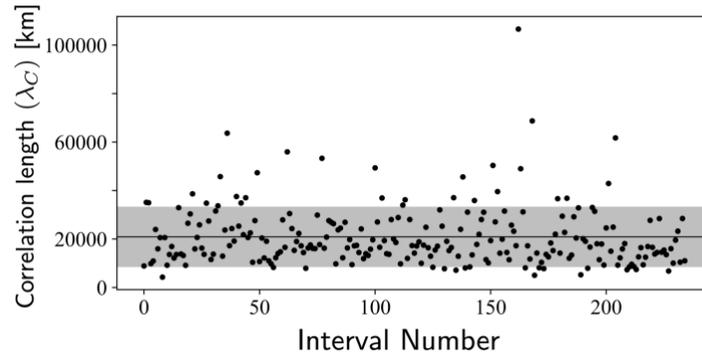

**Figure 4:** Correlation length estimates for the PSP first encounter. The error bars are smaller than the symbols, and therefore, cannot be seen here. The average and standard deviation are shown by the horizontal line and the shaded region.

To visualize the Taylor microscale and correlation length, we compute the turbulent power spectrum from the magnetic field data in the 15-minute period sampled in Figs. 1 and 2. Fig. 5 plots the magnetic-field power spectrum with the estimated Taylor microscale and correlation scale shown. We also show the ion-inertial length in the same figure.

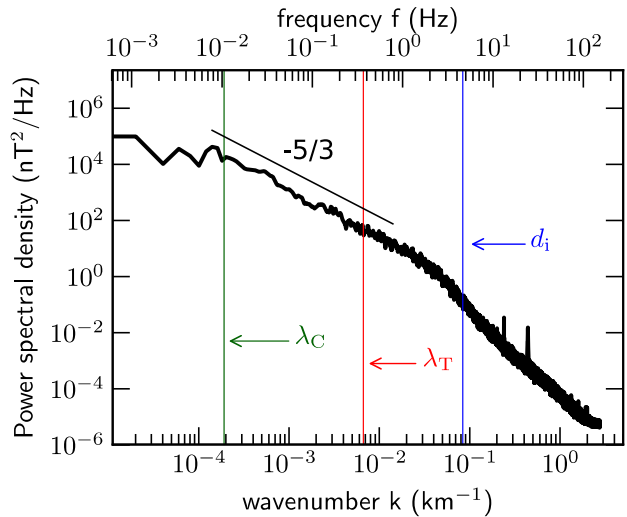

**Figure 5:** Magnetic power spectrum for the 15-min PSP data collected during 11:14:59 UTC to 11:29:59 UTC on 2018 November 6. Both frequency ($f$) and Taylor-shifted wavenumber ($k = 2\pi f/V_{SW}$) axes are shown. A Kolmogorov scaling ($\sim k^{-5/3}$) is shown for reference. Three characteristic length scales – the correlation scale ($\lambda_C$), Taylor microscale ($\lambda_T$), and ion-inertial length ($d_i$), are shown using vertical lines.

From Fig. 5, the Kolmogorov scaling holds roughly within the range between the correlation scale and the Taylor microscale, indicating an inertial range here. Further, the "break" or "knee" in the spectrum is in an intermediatory position between the Taylor microscale and the ion-inertial length, as is expected from hydrodynamic turbulence (Matthaeus et al. 2008).

Equation (7) produces the Reynolds number from the ratio of the correlation and Taylor scales. We find the uncertainty in the *Re*-values by propagating forward the uncertainties from the Taylor and correlation scales estimates. For example, the data of Figs. 1 and 2 give a Reynolds number of 1211 ± 32. We show the Reynolds number estimates for all the samples in Fig. 6. There appears a very weak time variation with early and late intervals having generally lower Reynolds numbers than the central intervals. This may be related to the increase of *Re*-values at smaller distances. Nevertheless, since the radial distance variation is small here, we do not explore dependence on radial distance here.

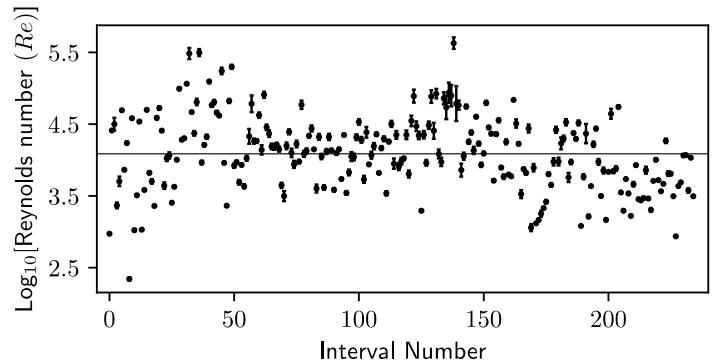

**Figure 6:** Estimated Reynolds number (in logarithmic scale) values. Note that we have taken the mean and standard deviation of the logarithm of *Re*-values because the Reynolds number values are widely distributed.

The Reynolds number values are distributed over a wide range, covering more than 3 orders of magnitude, and the logarithm of the *Re*-values produced a more Gaussian distribution. Therefore, following Ranquist 2020, we report the statistics on the logarithm and then convert back to the parameter (e.g., the logarithmic average of the *Re*-values is 4.09, which we convert to $10^{4.09} = 12241$). The average value and standard deviation obtained in this way are $\overline{Re} = 10^{4.09 \pm 0.52} = 12241^{+27570}_{-8526}$, as shown in Fig. 6.

Figure 7 shows the distributions of the evaluated Taylor microscales, correlation lengths, and Reynolds numbers from all the data in the 58.5-hour period. As before, the Reynolds number distribution is shown and the statistics are performed in logarithmic scale, while the other two use a linear scale.

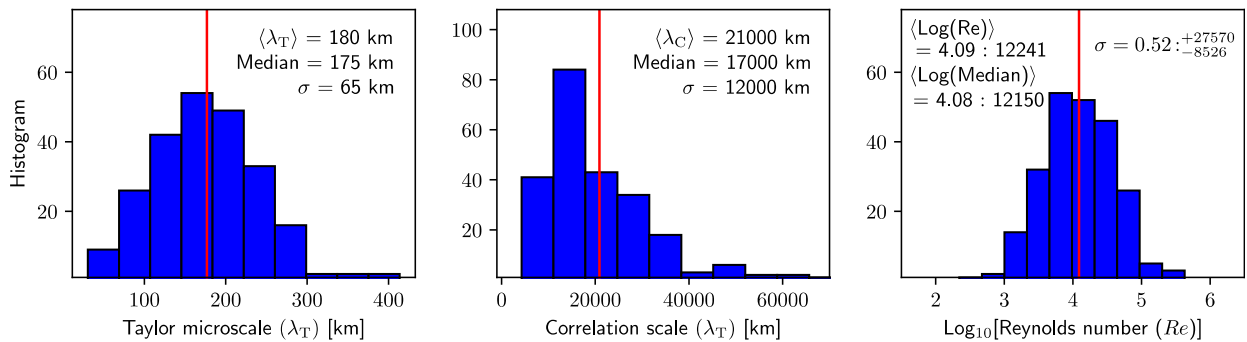

**Figure 7:** Distribution of (from left to right) Taylor scale, correlation length, and effective Reynolds number from PSP data. In each plot, the red vertical line shows the average value. The average value, median, and standard deviation are noted on each plot. Values to the right of a colon are the equivalent median, average, and standard deviation without a logarithm. Note that the Reynolds number histogram is plotted on a logarithmic scale in the x-axis while those for the correlation length and Taylor microscale are plotted on linear scales.

The Taylor microscale can also be used to estimate an effective viscosity/resistivity (or turbulent viscosity/resistivity), using the hydrodynamic turbulence formula

$$\nu = \frac{\epsilon}{2E} \frac{\lambda_T^2}{3}$$

where, we recall that $\epsilon$ is the average rate of dissipation. $E$ is the fluctuation energy per unit mass. Using the Politano-Pouquet third-order law, the cascade rate in PSP first encounter was estimated as ~$10^5$ W/kg (Andrés et al. 2021a, 2021b; Bandyopadhyay et al. 2020a; Hernández et al. 2021). Including the average Taylor microscale value of 180 km and $2E = \langle |\boldsymbol{b}|^2 \rangle \sim 10^4$ km$^2$/s$^2$, we obtain $\nu \sim 100$ km$^2$s$^{-1}$. This value is similar to the near-Earth estimates of solar wind viscosity (Bandyopadhyay et al. 2020b; Verma 1996; Verma et al. 1995), suggesting that the effective viscosity of the solar wind does not change very much on its journey from ~36 $R_s$ outwards.

## 5. Discussion and Summary

In this paper we calculate an average Taylor scale value of $\approx$ 180 km for the "young" solar wind at a heliocentric distance of $\approx$ 0.17 au (36 $R_S$) from PSP data. From the same data set, the average correlation length is found approximately $\approx$ 21000 km, which is close to those reported in previous studies using PSP data (Bandyopadhyay et al. 2020a; Chen et al. 2020; Parashar et al. 2020). Using the estimated Taylor scale and correlation length, we calculate the effective magnetic Reynolds number with average value of $\approx$ 12000. As far as we are aware, this is the first estimation of the Taylor microscale from solar wind data collected near the Sun.

The value of Taylor scale we obtained from the PSP data in the first solar encounter is much smaller than the previously reported value of ~ 2500 km at 1 AU (e.g., Weygand et al. 2011). This indicates that the onset of dissipative processes occurs at smaller scales in the near-Sun solar wind compared to that near Earth.

The Reynolds number values are also much smaller than the earlier published near-Earth value of ~ 230000 (e.g., Matthaeus et al. 2005). In fact, they are considerably smaller than even Helios 1 estimates of *Re*-values of ~ $10^6$ at $\approx$ 0.3 au (Cuesta et al. 2022). Extrapolating from Helios 1 and Voyager data, one would expect Reynolds number to increase as the instrument approaches the Sun (Cuesta et al. 2022). The Reynolds number instead drops sharply, indicating a transition point between ~0.17 au and ~0.3 au where the turbulence rapidly evolves. Our Reynolds number estimates are close to the ones reported recently by Cuesta et al. 2022, who used a different formalism independent of the Taylor microscale (see Eq. 6 and the text below it) to calculate the *Re*-values for PSP encounters 1-3. The abrupt decrease in Reynolds number close to the Sun is possibly related to the "aging" of the turbulence (Matthaeus et al. 1998; Ruiz et al. 2011), and may support the driving of turbulence by shear in the region right above the Alfvén zone (Bandyopadhyay & McComas 2021; Ruffolo et al. 2020). However, Ruiz et al. 2011 also showed that correlation scales measured parallel to the mean magnetic field are systematically smaller than those measured perpendicular to the mean magnetic field and that this difference grows stronger closer to the Sun. Since most PSP measurements of correlation scale are parallel to the mean magnetic field, the small correlation scale (and therefore Reynolds number) may be in part due to this systematic selection effect.

Finally, we note that although the obtained values of the Taylor microscale and Reynolds number are much smaller than those in the near-Earth solar wind, the estimated viscosity values are comparable. This result indicates that the evolving solar wind possibly maintains a fairly constant effective viscosity. Global solar wind models and space weather predicting models would benefit from these observations.

## Acknowledgements

We are deeply indebted to everyone that helped make the Parker Solar Probe (PSP) mission possible. This work was supported as a part of the PSP mission under contract NNN06AA01C. Parker Solar Probe was designed, built, and is now operated by the Johns Hopkins Applied Physics Laboratory as part of NASA's Living with a Star (LWS) program (contract NNN06AA01C). Support from the LWS management and technical team has played a critical role in the success of the Parker Solar Probe mission. This research was partially supported by the Parker Solar Probe Plus project through Princeton/IS☉IS subcontract SUB0000165 and in part by PSP GI grant 80NSSC21K1767 at Princeton University. The authors thank the FIELDS team (PI: Stuart D. Bale, UC Berkeley) and the SWEAP team (PI: Justin Kasper, BWX Technologies) for help with the PSP data. The authors also thank Sean Oughton and Bill Matthaeus for helpful discussions. All the data, used in this paper, are publicly available via the NASA Space Physics Data Facility (https://spdf.gsfc.nasa.gov/).

## References

Alexandrova, O., Saur, J., Lacombe, C., et al. 2009, Phys Rev Lett, 103 (American Physical Society), 165003
Andrés, N., Sahraoui, F., Hadid, L. Z., et al. 2021a, Astrophys J, 919 (American Astronomical Society), 19, https://doi.org/10.3847/1538-4357/ac0af5
Andrés, N., Sahraoui, F., Huang, Hadid, L. Z., & Galtier, S. 2021b, ArXiv E-Prints, arXiv:2112.13748, https://ui.adsabs.harvard.edu/abs/2021arXiv211213748A
Bale, S., Badman, S., Bonnell, J., et al. 2019, Nature (Nature Publishing Group), 1
Bale, S. D., Goetz, K., Harvey, P. R., et al. 2016, Space Sci Rev, 204, 49
Bandyopadhyay, R., Goldstein, M. L., Maruca, B. A., et al. 2020a, Astrophys J Suppl Ser, 246, 48
Bandyopadhyay, R., Matthaeus, W. H., Chasapis, A., et al. 2020b, Astrophys J, 899 (American Astronomical Society), 63
Bandyopadhyay, R., Matthaeus, W. H., Parashar, T. N., et al. 2020c, Astrophys J Suppl Ser, 246 (American Astronomical Society), 61
Bandyopadhyay, R., & McComas, D. J. 2021, Astrophys J, 923 (American Astronomical Society), 193
Batchelor, G. K. 1953, The theory of homogeneous turbulence (Cambridge university press)
Blackman, R. B., & Tukey, J. W. 1958, Bell Syst Tech J, 37, 185

Bowen, T. A., Bale, S. D., Bonnell, J. W., et al. 2020, J Geophys Res Space Phys, 125, e2020JA027813
Breech, B., Matthaeus, W. H., Minnie, J., et al. 2008, J Geophys Res Space Phys, 113, http://doi.org/10.1029/2007JA012711
Cartagena-Sanchez, C. A., Carlson, J. M., & Schaffner, D. A. 2022, Phys Plasmas, 29, 032305, https://doi.org/10.1063/5.0073207
Case, A. W., Kasper, J. C., Stevens, M. L., et al. 2020, Astrophys J Suppl Ser, 246 (American Astronomical Society), 43
Chen, C. H. K., Bale, S. D., Bonnell, J. W., et al. 2020, Astrophys J Suppl Ser, 246 (American Astronomical Society), 53
Chhiber, R., Usmanov, A. V., Matthaeus, W. H., & Goldstein, M. L. 2021, Astrophys J, 923 (American Astronomical Society), 89
Chuychai, P., Weygand, J. M., Matthaeus, W. H., et al. 2014, J Geophys Res Space Phys, 119, 4256
Coleman, P. J. 1968, Astrophys J, 153, 371
Cuesta, M. E., Parashar, T. N., Chhiber, R., & Matthaeus, W. H. 2022, Astrophys J Suppl Ser, 259, 23
Fox, N. J., Velli, M. C., Bale, S. D., et al. 2016, Space Sci Rev, 204, 7
Franz, M., & Harper, D. 2002, Planet Space Sci, 50, 217
Frisch, U. 1995, Turbulence (Cambridge, UK: cup)
Gurgiolo, C., Goldstein, M. L., Matthaeus, W. H., Viñas, A., & Fazakerley, A. N. 2013a, Ann Geophys, 31, 2063, https://angeo.copernicus.org/articles/31/2063/2013/
Gurgiolo, C., Goldstein, M. L., Matthaeus, W. H., Viñas, A., & Fazakerley, A. N. 2013b, Ann Geophys, 31, 2063
Hernández, C. S., Sorriso-Valvo, L., Bandyopadhyay, R., et al. 2021, Astrophys J Lett, 922 (American Astronomical Society), L11, https://doi.org/10.3847/2041-8213/ac36d1
Huang, S. Y., Sahraoui, F., Andrés, N., et al. 2021, Astrophys J Lett, 909 (American Astronomical Society), L7
Kasper, J., Bale, S., Belcher, J., et al. 2019, Nature (Nature Publishing Group), 1
Kasper, J. C., Abiad, R., Austin, G., et al. 2016, Space Sci Rev, 204, 131
Kolmogorov, A. N. 1941, Dokl Akad Nauk SSSR, 30, 301
Matthaeus, W. H., Dasso, S., Weygand, J. M., et al. 2005a, Phys Rev Lett, 95 (American Physical Society), 231101
Matthaeus, W. H., Dasso, S., Weygand, J. M., et al. 2005b, Phys Rev Lett, 95, 1
Matthaeus, W. H., & Goldstein, M. L. 1982, J Geophys Res, 87, 6011
Matthaeus, W. H., Smith, C. W., & Oughton, S. 1998, J Geophys Res Space Phys, 103, 6495
Matthaeus, W. H., Weygand, J. M., Chuychai, P., et al. 2008, apj, 678, L141
McComas, D., Christian, E., Cohen, C., et al. 2019, Nature (Nature Publishing Group), 1
Parashar, T. N., Cuesta, M., & Matthaeus, W. H. 2019, Astrophys J, 884 (American Astronomical Society), L57, https://doi.org/10.3847/2041-8213/ab4a82
Parashar, T. N., Goldstein, M. L., Maruca, B. A., et al. 2020, Astrophys J Suppl Ser, 246 (American Astronomical Society), 58
Parashar, T. N., Matthaeus, W. H., Shay, M. A., & Wan, M. 2015, Astrophys J, 811, 112
Pearson, B. R., Krogstad, P. A., & Water, W. van de. 2002, Phys Fluids, 14, 1288
Pearson, B. R., Yousef, T. A., Haugen, N. E. L., Brandenburg, A., & Krogstad, P. 2004, Phys Rev E, 70 (American Physical Society), 056301

Pope, S. B. 2000, Turbulent Flows (Cambridge: Cambridge University Press), https://www.cambridge.org/core/books/turbulent-flows/C58EFF59AF9B81AE6CFAC9ED16486B3A

Ranquist, D. A. 2020, Large Scale Structure of Jupiter's Magnetosphere and Magnetosheath (ProQuest Dissertations Publishing), http://ucblibraries.summon.serialssolutions.com/2.0.0/link/0/eLvHCXMwY2AwNtIz0EUrE1IMkoAVUZJhaqIhsAA0NEg1SjEzSDI3Nkm0SE5OSQVFfFCAhVegpVekSTB0cSFoaww0umGlJLjoTslPBo2a64MP5zIGEub2BYW6oHukQPOt0Es1mBlYQRdpg7pj7sgNIlj_ndfSHHQwObD9AOz_GGMUxODaxU2AIRPmENAR0kWJKfl6mSm58I3SaCc3UuRgQQYeF6QZeCEGptQ8EQYnH9B6cIVgYHwBSfCJsqVFqQr5aQpepQWgPcrqxQq-iel5qSX5xaBjCFIVEvNS4CKg4jxDlEHZzTXE2UMX5qZ4aCotjkc4yFiMgSUvPy9VgkHBPNUg1SIpOSXRDBg-RmkpiaamCmjqVxkpmFJIMMPpOk8EtLM3AZgXqs4EEMGQYWoJdSZRl4SpOT4MMDcgzMBr6BYDJIDhx9AA40sLU

Ruffolo, D., Matthaeus, W. H., Chhiber, R., et al. 2020, Astrophys J, 902 (American Astronomical Society), 94

Ruiz, M. E., Dasso, S., Matthaeus, W. H., Marsch, E., & Weygand, J. M. 2011, J Geophys Res Space Phys, 116, https://agupubs.onlinelibrary.wiley.com/doi/abs/10.1029/2011JA016697

Sahraoui, F., Goldstein, M. L., Robert, P., & Khotyaintsev, Yu. V. 2009, Phys Rev Lett, 102 (American Physical Society), 231102, https://doi.org/10.1103/PhysRevLett.102.231102

Smith, C. W., Matthaeus, W. H., Zank, G. P., et al. 2001, J Geophys Res Space Phys, 106, 8253

Sorriso-Valvo, L., Marino, R., Carbone, V., et al. 2007, Phys Rev Lett, 99 (American Physical Society), 115001, https://link.aps.org/doi/10.1103/PhysRevLett.99.115001

Taylor, G. I. 1938, Proc R Soc Lond Ser A, 164, 476

Tennekes, H., & Lumley, J. L. 1972, A First Course in Turbulence (The MIT Press), https://doi.org/10.7551/mitpress/3014.001.0001

Tessein, J. A., Matthaeus, W. H., Wan, M., et al. 2013, Astrophys J Lett, 776, L8

Usmanov, A. V., Matthaeus, W. H., Goldstein, M. L., & Chhiber, R. 2018, Astrophys J, 865 (American Astronomical Society), 25

Vech, D., Kasper, J. C., Klein, K. G., et al. 2020, Astrophys J Suppl Ser, 246 (American Astronomical Society), 52, https://doi.org/10.3847/1538-4365/ab60a2

Verma, M. K. 1996, J Geophys Res Space Phys, 101, 27543, https://agupubs.onlinelibrary.wiley.com/doi/abs/10.1029/96JA02324

Verma, M. K., Roberts, D. A., & Goldstein, M. L. 1995, J Geophys Res Space Phys, 100, 19839, https://agupubs.onlinelibrary.wiley.com/doi/abs/10.1029/95JA01216

Weygand, J. M., Matthaeus, W. H., Dasso, S., et al. 2009, J Geophys Res Space Phys, 114, https://agupubs.onlinelibrary.wiley.com/doi/abs/10.1029/2008JA013766

Weygand, J. M., Matthaeus, W. H., Dasso, S., & Kivelson, M. G. 2011, J Geophys Res Space Phys, 116, https://agupubs.onlinelibrary.wiley.com/doi/abs/10.1029/2011JA016621

Weygand, J. M., Matthaeus, W. H., Passo, S., Kivelson, M. G., & Walker, R. J. 2007, J Geophys Res Space Phys, 112, 1

Zhou, G., & He, H.-Q. 2021, Astrophys J Lett, 911 (American Astronomical Society), L2

Zhou, G., He, H.-Q., & Wan, W. 2020, Astrophys J Lett, 899 (American Astronomical Society), L32